\documentclass[%
superscriptaddress,
 amsmath,amssymb,
 aps,
]{revtex4-1}

\usepackage{graphicx}% Include figure files
\usepackage{dcolumn}% Align table columns on decimal point
\usepackage{bm}% bold math
\usepackage{booktabs}

\begin{document}
 
\preprint{APS/123-QED}

\title{Investigation of $\alpha$-induced reactions on Sb isotopes \\ relevant to the astrophysical $\gamma$-process}

\author{Z. Korkulu}
\email{korkulu@ribf.riken.jp}
\altaffiliation{Present address: RIKEN Nishina Center, 2-1 Hirosawa, Wako, Saitama, 351-0198, Japan}
\affiliation{Department of Physics, Kocaeli University, Umuttepe Kocaeli, 41350, Turkey}
\affiliation{MTA Atomki, P.O. Box 51, H-4001 Debrecen, Hungary} 
\author{N. {\"{O}}zkan}
\affiliation{Department of Physics, Kocaeli University, Umuttepe Kocaeli, 41350, Turkey}
\author{G. G. Kiss} 
\affiliation{MTA Atomki, P.O. Box 51, H-4001 Debrecen, Hungary}
\author{T. Sz{\"{u}}cs}
\affiliation{MTA Atomki, P.O. Box 51, H-4001 Debrecen, Hungary}
\author{Gy. Gy{\"{u}}rky}
\affiliation{MTA Atomki, P.O. Box 51, H-4001 Debrecen, Hungary}
\author{Zs. F{\"{u}}l{\"{o}}p}
\affiliation{MTA Atomki, P.O. Box 51, H-4001 Debrecen, Hungary}
\author{R.~T. G{\"{u}}ray} 
\affiliation{Department of Physics, Kocaeli University, Umuttepe Kocaeli, 41350, Turkey}
\author{Z. Hal\'asz} 
\affiliation{MTA Atomki, P.O. Box 51, H-4001 Debrecen, Hungary}
\author{T. Rauscher}
\affiliation{Department of Physics, University of Basel, 4056, Switzerland}
\affiliation{Centre for Astrophysics Research, University of Hertfordshire, Hatfield AL10 9AB, United Kingdom}
\author{E. Somorjai}
\affiliation{MTA Atomki, P.O. Box 51, H-4001 Debrecen, Hungary}
\author{Zs. T\"{o}r\"{o}k}
\affiliation{MTA Atomki, P.O. Box 51, H-4001 Debrecen, Hungary}
\author{C. Yal{\c{c}}{\i}n}
\affiliation{Department of Physics, Kocaeli University, Umuttepe Kocaeli, 41350, Turkey}

\date{\today}% It is always \today, today,

\begin{abstract}
\begin{description}

\item[Background]
The reaction rates used in $\gamma$-process nucleosynthesis network calculations are mostly derived from theoretical, statistical model cross sections. Experimental data is scarce for charged particle reactions at astrophysical, low energies. Where experimental ($\alpha$,$\gamma$) data exists, it is often strongly overestimated by Hauser-Feshbach statistical model calculations. Further experimental $\alpha$-capture cross sections in the intermediate and heavy mass region are necessary to test theoretical models and to gain understanding of heavy element nucleosynthesis in the astrophysical $\gamma$-process.

\item[Purpose]
The aim of the present work is to measure the $^{121}$Sb($\alpha,\gamma$)$^{125}$I, $^{121}$Sb($\alpha$,n)$^{124}$I, and $^{123}$Sb($\alpha$,n)$^{126}$I reaction cross sections. These measurements are important tests of astrophysical reaction rate predictions and extend the experimental database required for an improved understanding of p-isotope production.

\item[Method]
The $\alpha$-induced reactions on natural and enriched antimony targets were investigated using the activation technique. The ($\alpha$,$\gamma$) cross sections of $^{121}$Sb were measured and are reported for first time. To determine the cross section of the $^{121}$Sb($\alpha$,$\gamma$)$^{125}$I, $^{121}$Sb($\alpha$,n)$^{124}$I, and $^{123}$Sb($\alpha$,n)$^{126}$I reactions, the yields of $\gamma$-rays following the $\beta$-decay of the reaction products were measured. For the measurement of the lowest cross sections, the characteristic X-rays were counted with a LEPS (Low Energy Photon Spectrometer) detector.

\item[Results]
The cross section of the $^{121}$Sb($\alpha$,$\gamma$)$^{125}$I, $^{121}$Sb($\alpha$,n)$^{124}$I and $^{123}$Sb($\alpha$,n)$^{126}$I reactions were measured with high precision in an energy range between 9.74 MeV to 15.48 MeV, close to the astrophysically relevant energy window. The results are compared with the predictions of statistical model calculations. The ($\alpha$,n) data show that the $\alpha$ widths are predicted well for these reactions. The ($\alpha$,$\gamma$) results are overestimated by the calculations but this is due to the applied neutron- and $\gamma$ widths.

\item[Conclusions]
Relevant for the astrophysical reaction rate is the $\alpha$ width used in the calculations. While for other reactions the $\alpha$ widths seem to have been overestimated and their energy dependence was not described well in the measured energy range, this is not the case for the reactions studied here. The result is consistent with the proposal that additional reaction channels, such as Coulomb excitation, may have led to the discrepancies found in other reactions.

\end{description}
\end{abstract}
%\begin{description}
%\pacs{25.40.Lw, 26.30.-k, 29.30.Kv}
%\pacs{ {25.40.Lw} {Radiative capture}
%      {26.30.-k} {Nucleosynthesis in novae, supernovae, and other explosive environments}
%      {29.30.Kv} {X- and γ-ray spectroscopy}
%} %% PACS, the Physics and Astronomy
%\end{description}

%\end{document}

%\begin{document}  

\maketitle
%\tableofcontents

\section{Introduction}
\label{sec:intro}
The production of nuclei up to iron occurs in stars during different stellar burning phases. Nuclei heavier than Fe are synthesized mostly by the neutron capture reactions that typify the s- and r-processes \cite{Burbidge,arnould07}. The s-process (slow neutron capture process) is the mechanism for the formation of about half of the nuclides between Fe and Bi and proceeds through successive neutron capture reactions and subsequent \mbox{$\beta$$^{-}$ decays}. The r-process (rapid neutron capture process) is responsible for the synthesis of the other half of heavy nuclei including isotopes above Bi. 
Recent observations and simulations have suggested that, in addition to the well-known s- and r-processes, there may be an intermediate mode of neutron capture nucleosynthesis, the so-called i-process \cite{J.Cowan}.

Since the build-up of nuclides in the s- and r-process either follows the valley of stability or populates the neutron-rich side, about 35 proton-rich isotopes cannot be produced by these neutron capture mechanisms. These so-called p-nuclei are typically 10 to 100 times less abundant than the s- and r-nuclei, and can be produced by the photodisintegration of pre-existing intermediate and heavy nuclei \cite{Raus13}. For the production of neutron-deficient p-nuclei, ($\gamma$,n) photodisintegration reactions are initiated on s- or r-process seeds. Since the neutron separation energy increases after several neutron emissions, ($\gamma$,$\alpha$) and ($\gamma$,p) reactions start to compete with ($\gamma$,n) reactions and the reaction path is deflected towards the lower mass region. While ($\gamma$,$\alpha$) reactions are mainly important for the abundance of medium and heavy mass p-nuclei, the ($\gamma$,p) reactions are important for the production of the lower mass p-nuclei \cite{Raus06, W.Rapp06}. The process mentioned above is called the $\gamma$-process and requires the sufficiently high temperatures ($2 - 3$ GK) achieved in pre-explosive or explosive O/Ne burning of massive stars \cite{Arnould,Woosley78}. An alternative production scenario is a subclass of type Ia supernova explosion of sub-Chandrasekhar-mass white dwarfs (mainly composed of C and O) \cite{Trav10}.

Although the $\gamma$-process is thought to be the main mechanism contributing to the synthesis of p-nuclei, different processes also seem to contribute. These processes include the \mbox{rp-process} \cite{Schatz}, the $\nu$p-process \cite{Frohlich}, the pn-process \cite{Goriely} and the $\nu$-process \cite{Woosley}. The rp-process,  the $\nu$p-process and the pn-process can contribute to the nucleosynthesis of light p-nuclei, while the very rare p-nuclei $^{138}$La and $^{180}$Ta$^{m}$ probably receive a large contribution from the \linebreak $\nu$-process.  

Measurements of nuclear reaction cross sections are crucial for $\gamma$-process models since modeling the $\gamma$-process requires the knowledge of thousands of photodisintegration cross sections, which are based on mostly untested at sub-Coulomb energies theoretical calculations obtained from the statistical Hauser-Feshbach approach. Due to the effect of thermal nuclear excitation in the hot stellar plasma, it is favorable to study the inverse charged particle capture reactions instead of the photodisintegrations directly. The reaction rates can then be obtained by applying the detailed balance theorem \cite{MohrP,Raus09}. The available experimental cross sections of the (p,$\gamma$) reactions generally agree with the statistical model predictions within about a factor of two \cite{Raus13}, while in the case of reactions involving \linebreak $\alpha$-particles much larger deviations are found. This has spurred a number of experimental investigations of low-energy $\alpha$ capture on nuclei in the mass range of $p$ nuclei (see the references in \cite{Raus13} and, more recently, \cite{har13,qui14, ornelas,simon15,quinn15}). So far, several ($\alpha$,$\gamma$) cross sections around \mbox{A $\approx$ 100} have been studied via the activation method \cite{Ozkan02,Gy06,Ozkan07,Bas07,Rapp2006,Cata08,Yalcin09,Gy10,Kiss11,Fili11,Kiss12,halasz12,net13,kis14,yal15,Net15,halasz16,tamas18}. Often, the obtained ($\alpha$,$\gamma$) cross sections are considerably lower than the model predictions. The calculations also poorly reproduce the energy dependence of the cross sections. This could potentially be one of the reasons that self-consistent models of the $\gamma$-process fail to synthesize the p-nuclei in the required amount, especially in the mass regions \mbox{A $<$ 124} and \linebreak 150 $\leq$ A $\leq$ 165 \cite{Raus02}. It is therefore necessary to test and improve model calculations through the collection of experimental data in the relevant mass and energy range.

In order to fulfill this goal, and to extend the available experimental database relevant for the $\gamma$-process, a systematic investigation of $\alpha$-induced reactions is being carried out at the Institute for Nuclear Research in Debrecen, Hungary (Atomki) \cite{atomki}. In this paper we present the experimental technique and the results of alpha induced reaction cross section measurements on Sb isotopes. For the first time, $^{121}$Sb($\alpha$,$\gamma$)$^{125}$I cross sections have been measured in the center of mass energy range between 9.74 MeV and 13.54 MeV. These energies are close to the astrophysically relevant energy range (the so-called Gamow window), which covers 6.15 MeV -- 8.68 MeV at a typical $\gamma$-process temperature of T\,=\,3\,GK \cite{Raus10}. Although the ($\alpha$,n) reactions have no direct astrophysical relevance, the analysis of previous experiments shows that the comparison of measured and calculated ($\alpha$,n) cross sections can provide useful insights regarding the selection of input parameters for the calculations (see Sec.\,\ref{sec:comparison}). Therefore, along with the ($\alpha$,$\gamma$) measurement, the cross section of the $^{121}$Sb($\alpha$,n)$^{124}$I and $^{123}$Sb($\alpha$,n)$^{126}$I reactions have also been measured.

The investigated reactions are discussed in detail in Sec.\,\ref{sec:invesreac}, the experimental procedure is described in Sec.\,\ref{sec:exp} and the results with a comparison to statistical model calculations are given in Sec.\,\ref{sec:results}. Finally, in Sec.\,\ref{sec:conclusions} a summary and conclusions are provided.

\section{Investigated Reactions}
\label{sec:invesreac}

The element antimony has two stable isotopes: $^{121}$Sb and $^{123}$Sb with natural abundances of 57.21\% and 42.79\%, respectively. 
The cross sections of the $^{121}$Sb($\alpha$,$\gamma$)$^{125}$I, $^{121}$Sb($\alpha$,n)$^{124}$I and $^{123}$Sb($\alpha$,n)$^{126}$I reactions were measured with the activation method, since all the reaction products are radioactive and have at least one strong gamma-ray released after the decay. 
The decay parameters of the reaction products are listed in Table\,\ref{tab:decaypar}. 
$^{125}$I has a half-life of 59.40 days and its electron capture decay is followed by the emission of a 35.49 keV $\gamma$-ray. The detection of this radiation was used for the cross section measurement of $^{121}$Sb($\alpha$,$\gamma$)$^{125}$I. For the sake of low cross section measurements, the detection of the characteristic X-rays following the electron capture decay was also used since these X-rays have higher relative intensities than the $\gamma$-ray.
\raggedbottom

 In order to measure both ($\alpha$,n) cross sections in a single activation, natural isotopic composition targets were used. In the case of both ($\alpha$,n) reactions, the relative intensities of the $\gamma$-rays following the decay of the reaction products are high enough. Therefore, the ($\alpha$,n) cross sections were measured via $\gamma$-counting only.

Using natural isotopic composition targets has two disadvantages. Since the product of the $^{121}$Sb($\alpha$,$\gamma$)$^{125}$I reaction is the same as that of $^{123}$Sb($\alpha$,2n)$^{125}$I, the $^{121}$Sb($\alpha$,$\gamma$)$^{125}$I cross section can be determined with activation only below the ($\alpha$,2n) threshold located at 14.6\,MeV. This is not a serious limitation in our case however, since we require data in the low energy region. The other disadvantage is related to the X-ray detection method. The same energy X-rays are emitted from the reactions on both Sb isotopes, which thus cannot be distinguished. Especially at low energies where the X-ray counting was necessary, therefore, highly enriched (99.59$\%$) $^{121}$Sb targets were used. In these activations the $^{123}$Sb($\alpha$,n)$^{126}$I cross section could, of course, not be determined.

Considering that the ($\alpha$,n) cross sections are typically much higher than ($\alpha$,$\gamma$) cross sections, the low $^{123}$Sb content of the enriched targets can still affect the ($\alpha$,$\gamma$) cross section measurement. This was avoided by capitalizing on the different half-lives of the reaction products. The half-lives of $^{125}$I, $^{124}$I and $^{126}$I are 59.4 days, 4.2 days and 12.9 days, respectively. Exploiting the long half-life of $^{125}$I, characteristic X-ray counting was carried out after a long (about 10 to 14 weeks after the irradiations) waiting time, when the $^{126}$I and $^{124}$I activities in the irradiated enriched targets had decreased to low values (always less than 1.7\%).

No cross section measurement for the $^{121}$Sb($\alpha$,$\gamma$)$^{125}$I reaction has been carried out so far. Previous results for $^{121}$Sb($\alpha$,n)$^{124}$I and $^{123}$Sb($\alpha$,n)$^{126}$I reactions are, on the other hand, available in the literature
\cite{Uddin10,Tark09,Sin06,Has06,Is90}. The results of those measurements are included in Sec.\,\ref{sec:results} for comparison.

\begin{table*}
\caption{\label{tab:decaypar} Decay parameters of the $^{121}$Sb($\alpha$,$\gamma$)$^{125}$I, $^{121}$Sb($\alpha$,n)$^{124}$I and $^{123}$Sb($\alpha$,n)$^{126}$I reaction products taken from the literature \cite{NUDAT}.}
\setlength{\extrarowheight}{0.1cm}
\begin{ruledtabular}
\begin{tabular}{ccccc}
\parbox[t]{2.0cm}{\centering{Product nucleus}} &
\parbox[t]{1.5cm}{\centering{Decay mode}} &
\parbox[t]{1.5cm}{\centering{Half-life (d) }} &
\parbox[t]{3.0cm}{\centering{X- and $\gamma$-ray energy [keV] }} &
\parbox[t]{2.5cm}{\centering{Relative intensity per decay ($\%$)}}
\\

\hline

$^{125}$I& $\varepsilon$ 100$\%$ & 59.40 $\pm$ 0.01 & 27.202 (K$_{\alpha 2}$) & 39.6 $\pm$ 1.1 \\

 &  & & 27.472 (K$_{\alpha 1}$) & 73.1 $\pm$ 1.9 \\

 &  & & 35.49 & 6.68 $\pm$ 0.13 \\

$^{124}$I &$\varepsilon$ 100$\%$ & 4.1760 $\pm$ 0.0003 & 27.202 (K$_{\alpha 2}$) & 16.6 $\pm$ 0.8 \\
   &   &     &       27.472 (K$_{\alpha 1}$) & 30.6 $\pm$ 1.4 \\

   &   &     &       602.73 & 62.9 $\pm$ 0.7  \\

   &   &     &       722.78 & 10.36 $\pm$ 0.12  \\

   &   &     &       1690.96 & 11.15 $\pm$ 0.17 \\

$^{126}$I &$\varepsilon$ 52.7$\%$ & 12.93 $\pm$ 0.05 & 27.202 (K$_{\alpha 2}$) & 11.1 $\pm$ 0.4 \\

   &   &     &       27.472 (K$_{\alpha 1}$) & 20.4 $\pm$ 0.6 \\

   &   &     &       666.33 & 32.9 $\pm$ 0.7  \\

   &   &     &       753.82 & 4.15 $\pm$ 0.09  \\

   & $\beta$$^{-}$ 47.3$\%$  &     &  29.461 (K$_{\alpha 2}$) & 0.146 $\pm$ 0.006 \\

   &   &     &       29.782 (K$_{\alpha 1}$) & 0.269 $\pm$ 0.011 \\

   &   &     &       388.63 & 35.6 $\pm$ 0.6  \\

   &   &     &       491.24 & 2.88 $\pm$ 0.05  \\

\end{tabular}
\end{ruledtabular}
\end{table*}

\section{Experimental technique}
\label{sec:exp}

\subsection{Target preparation and irradiation}
\begin{figure}
 \centering
\resizebox{0.8\textwidth}{!}{
\includegraphics[-10,10] [800,550]{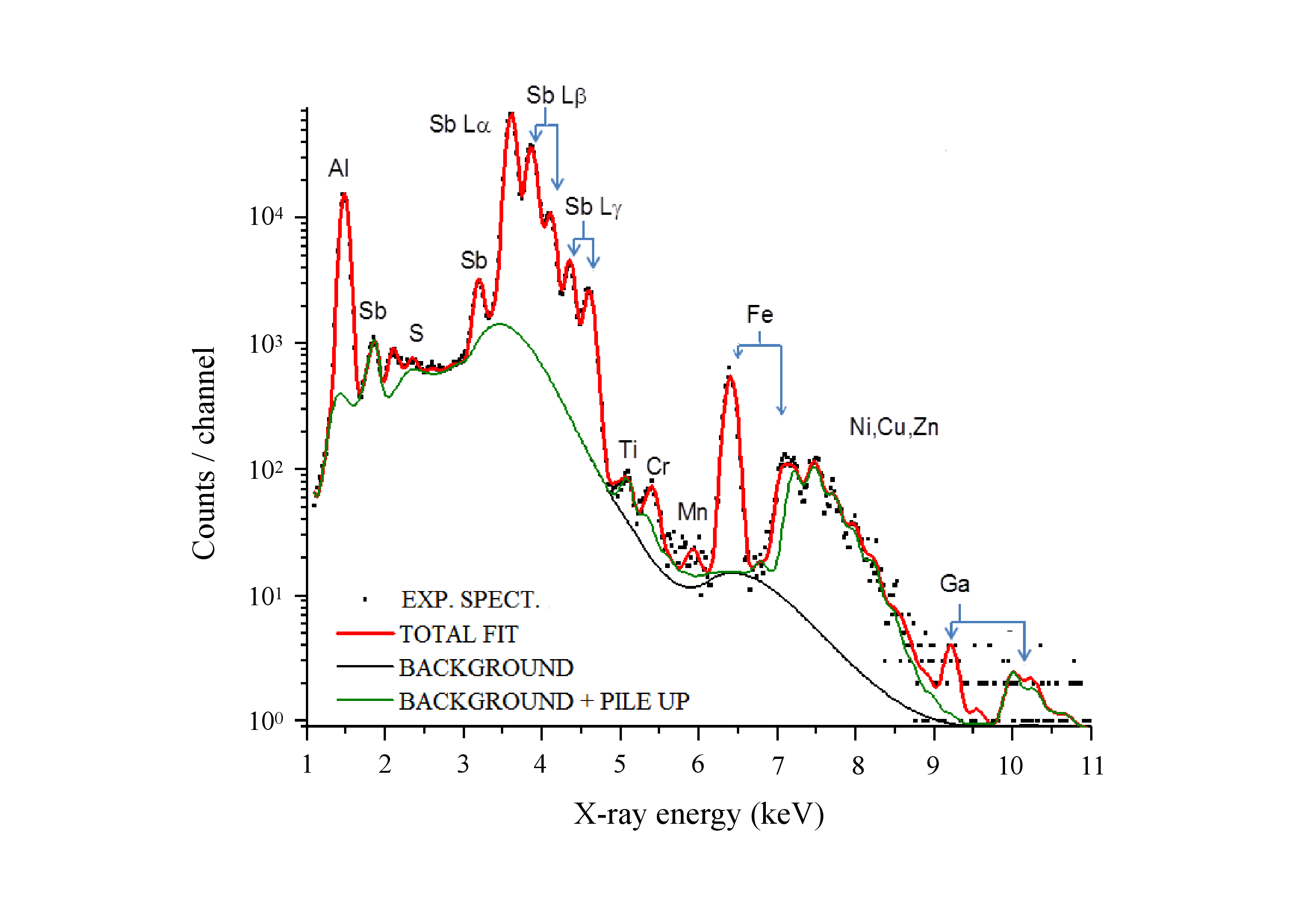}}
\caption{\label{fig:pixe} (Color online) PIXE spectrum measured with a proton beam of 2 MeV energy. The peaks used for the analysis are marked. Peaks belonging to impurities in target and the backing are also indicated.}
\end{figure}

The targets were produced by vacuum evaporation of natural Sb and enriched $^{121}$Sb (99.59$\%$) onto high purity thin (1.8 and 2.5 $\mu$m) Al foils of 12 mm diameter. The enriched metallic powder of $^{121}$Sb was obtained from the company TRACE (Certificate no: $\#$197-2a) \cite{trace}. The target thicknesses were determined in three ways: 1) weighing, 2) Rutherford backscattering (RBS) technique, and 3) the proton induced X-ray emission (PIXE) method. The weight of each foil was measured before and after the evaporation. The target thicknesses were then calculated from the weight difference. 
The uncertainty of the target thicknesses was found to be 7\% by taking into account the precision of the weight measurements (better than  5 $\mu$g) before and after the evaporation and the uncertainty of the target area determination.
For the RBS measurements the antimony targets were irradiated with an $\alpha$-beam of 4.7 MeV. The obtained RBS spectra were analyzed with the SIMNRA software, version 6.06 \cite{sim}. 
The uncertainty of the number of target atoms was found to be 5$\%$ for the RBS method, due to the uncertainty of the stopping power as a systematic error and the statistical error from the fit of the RBS spectra. 
The PIXE measurements were carried out using the PIXE setup of the MTA Atomki installed on the left 45$^{\circ}$ beamline of the 5 MV Van de Graaff accelerator \cite{kert}. The targets were irradiated with a proton beam of 2 MeV and the beam spot had a diameter of 5 mm. The total collected charge in the case of each target was about 1 $\mu$C. The obtained spectrum was fitted using the GUPIXWIN program code \cite{gupix}. A typical PIXE spectrum is shown in Fig.\,\ref{fig:pixe} and the peaks used for the analysis are marked. 
The fitting error is below 1$\%$ for each spectrum. The final uncertainty of 4$\%$ includes the systematic uncertainties concerning the geometry of the setup and the accuracy of the charge measurement. 
%The Sb targets thicknesses were determined with uncertainties of 4$\%$ from the PIXE method. 
%The thickness and uniformity of targets were found to be consistent using the PIXE and RBS methods.

\begin{figure}
 \centering
\resizebox{0.85\textwidth}{!}{
\includegraphics[-10,10] [700,450]{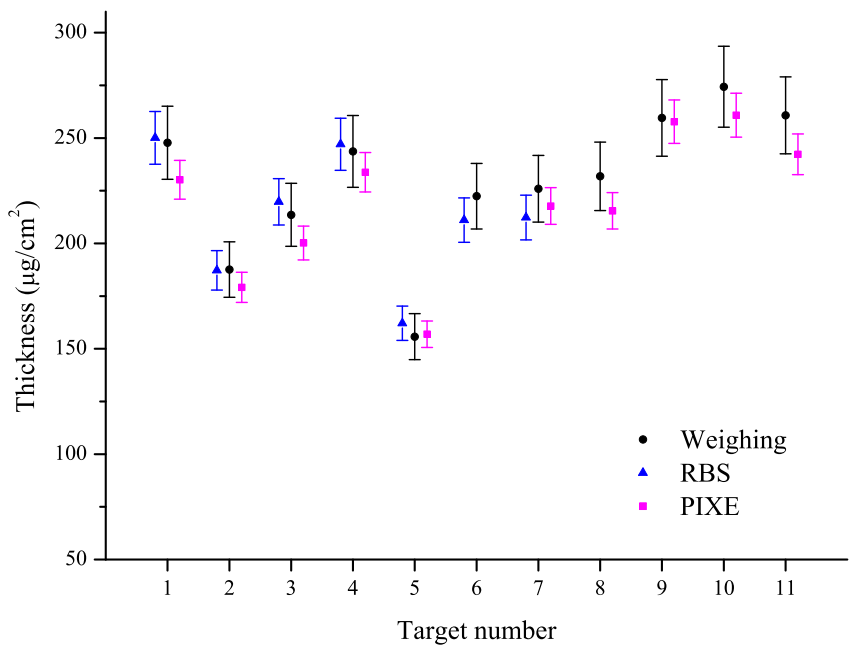}}
\caption{\label{fig:target} (Color online) Thicknesses of the Sb targets determined by the three methods. The data points for each target are segregated for better visualization.}
\end{figure}

The enriched targets were prepared with an areal density varying between 214 $\mu$g/cm$^{2}$ and 265 $\mu$g/cm$^{2}$ corresponding to an areal number densities of 1.1\,$\times$\,10$^{18}$ and 1.3\,$\times$\,10$^{18}$ $^{121}$Sb atoms/cm$^{2}$. The natural target thicknesses were between 159 $\mu$g/cm$^{2}$ and 241 $\mu$g/cm$^{2}$ (7.9\,$\times$\,10$^{17}$ -- 1.2\,$\times$\,10$^{18}$ Sb atoms/cm$^{2}$). 
The results of the three independent target-thickness determinations were in good agreement and the weighted average of results from the three methods has been adopted as the final result of the target thickness with an uncertainty of 4$\%$.
The results of the independent thickness measurements are presented in Fig.\,\ref{fig:target}. 
In some cases not all three methods were applied because of technical reasons. 

The target activations were carried out at the MGC cyclotron accelerator of Atomki. The targets were irradiated at 11 different alpha beam energies between 10.09 MeV and 16.00 MeV. For the $^{121}$Sb($\alpha$,$\gamma$)$^{125}$I reaction the energy range of 10.09 MeV to 14.00 MeV was covered in 0.5 MeV steps.

At certain energies the cyclotron cannot provide an $\alpha$-beam with sufficient intensity. In these cases a higher energy beam was used and an Al foil was placed in front of the target. Degrader thicknesses, determined via alpha energy loss measurements to a precision of 7\% using an $^{241}$Am $\alpha$-source and an ORTEC SOLOIST alpha spectrometer, were between 2.04 -- 6.62 $\mu$m.

\begin{figure}
 \centering
 \resizebox{0.7\textwidth}{!}{
 \includegraphics[-20,20] [1800,1100]{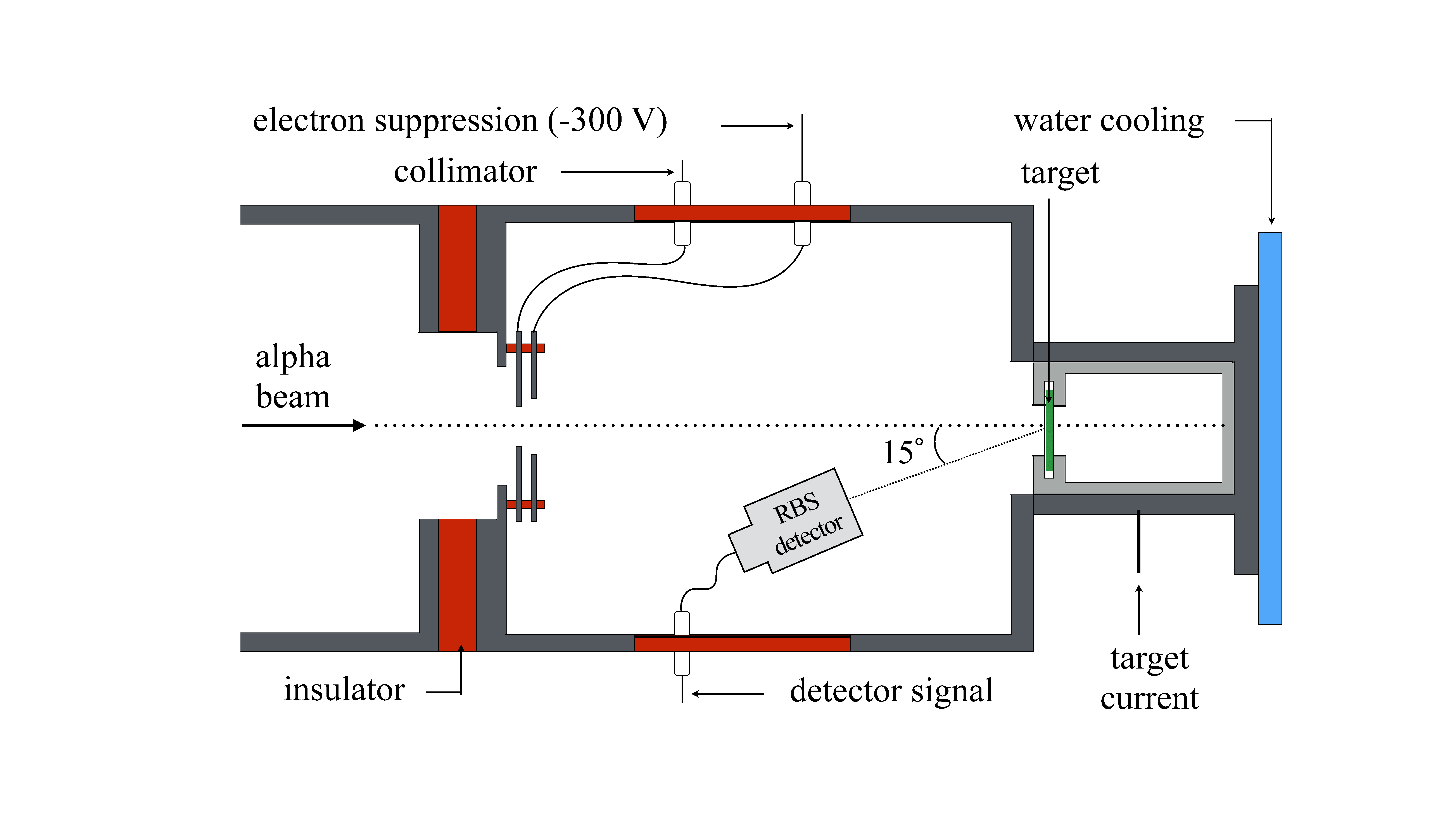}}
 \caption{\label{fig:chamber} (Color online) A drawing of the target chamber used for the irradiations.}
 \end{figure}

The schematic view of the irradiation chamber, which also serves as a Faraday cup, is shown in Fig.\,\ref{fig:chamber}. Secondary electrons were suppressed by --300 V at the entrance of the irradiation chamber. The duration of the irradiations was about 24 hours and the He$^{++}$-beam current was restricted to 1 $\mu$A. Before the irradiations a run was performed with a natural Sb target to determine the maximally allowed beam current that did not result in any deterioration of the target. This test showed that there was no target deterioration up to a current of 1 $\mu$A. The target stability was monitored during the irradiation by detecting the backscattered $\alpha$-particles with an ion implanted Si detector built into the irradiation chamber at 165$^{\circ}$ relative to the beam direction (see Fig.\,\ref{fig:chamber}). The number of $\alpha$ particles impinging on the targets were derived from the measurement of the collected charge, recorded in one minute intervals, by using a multi-channel scaler to monitor the changes of beam intensity.

\begin{table}
\caption{\label{tab:meas} The energy of the beam (E$_{\alpha}$), the waiting time after the end of irradiation (t$_{w}$), the counting time (t$_{c}$), and the studied reaction channels are given.}
\setlength{\extrarowheight}{0.2cm}
\begin{ruledtabular}
\begin{tabular}{cccccccccc} 
\parbox[t]{1.1cm}{\centering{E$_{\alpha}$\newline[MeV]}} &
\parbox[t]{1.1cm}{\centering{t$_{w1}$\newline[d]}} &
\parbox[t]{1.1cm}{\centering{t$_{c1}$\newline[d]}} &
\parbox[t]{1.1cm}{\centering{Studied \newline reaction}}\footnote[1]{measured with the HPGe detector.} &
\parbox[t]{1.1cm}{\centering{t$_{w2}$\newline[d]}} &
\parbox[t]{1.1cm}{\centering{t$_{c2}$\newline[d]}} &
\parbox[t]{1.1cm}{\centering{Studied \newline reaction}}\footnote[2]{measured with the LEPS detector.}&
\parbox[t]{1.1cm}{\centering{t$_{w3}$\newline[d]}} &
\parbox[t]{1.1cm}{\centering{t$_{c3}$\newline[d]}} &
\parbox[t]{1.1cm}{\centering{Studied \newline reaction}}\footnotemark[2] \\

\hline

10.09\footnote[3]{measured with an energy degrader foil.}$^{,}$\footnote[4]{enriched targets; the others are natural targets.} & 0.55 &  0.52 &  $^{121}$Sb($\alpha$,n) &   &  &    & 83 & 3 &   \vtop{\hbox{\strut $^{121}$Sb($\alpha$,$\gamma$)} \hbox{\strut via X-rays}}    \\

 10.43\footnotemark[3]$^{,}$\footnotemark[4] & 0.23  & 0.33 &  $^{121}$Sb($\alpha$,n)  &  &   &   & 73 & 3 &  \vtop{\hbox{\strut $^{121}$Sb($\alpha$,$\gamma$)} \hbox{\strut via X-rays}}   \\

 11.00\footnotemark[4] & 0.33 & 0.04 & $^{121}$Sb($\alpha$,n)   &   &  &   & 104 & 1.8 &   \vtop{\hbox{\strut $^{121}$Sb($\alpha$,$\gamma$)} \hbox{\strut via X-rays}}   \\

11.50\footnotemark[4] & 0.57  & 0.44 &   $^{121}$Sb($\alpha$,n)    & 31  &  10 &   \vtop{\hbox{\strut $^{121}$Sb($\alpha$,$\gamma$)} \hbox{\strut via $\gamma$-ray}}     &  106 & 1.7 &   \vtop{\hbox{\strut $^{121}$Sb($\alpha$,$\gamma$)} \hbox{\strut via X-rays}}    \\

12.00 & 0.08  & 0.08  &  \vtop{\hbox{\strut $^{121}$Sb($\alpha$,n)} \hbox{\strut $^{123}$Sb($\alpha$,n)}}   &  48  & 8 &   \vtop{\hbox{\strut $^{121}$Sb($\alpha$,$\gamma$)} \hbox{\strut via $\gamma$-ray}}  \\

12.53\footnotemark[3]$^{,}$\footnotemark[4] & 0.61  & 0.25   &   $^{121}$Sb($\alpha$,n)      & 32  & 3  &  \vtop{\hbox{\strut $^{121}$Sb($\alpha$,$\gamma$)} \hbox{\strut via $\gamma$-ray}}   &   101 & 0.3 &    \vtop{\hbox{\strut $^{121}$Sb($\alpha$,$\gamma$)} \hbox{\strut via X-rays}} \\

13.07\footnotemark[3] & 2.81 & 0.83  &      \vtop{\hbox{\strut $^{121}$Sb($\alpha$,n)} \hbox{\strut $^{123}$Sb($\alpha$,n)}}  & 60 & 5 &  \vtop{\hbox{\strut $^{121}$Sb($\alpha$,$\gamma$)} \hbox{\strut via $\gamma$-ray}}    \\

 13.50\footnotemark[4] & 0.97  & 0.03  &  $^{121}$Sb($\alpha$,n)      & 52  &  0.8 &   \vtop{\hbox{\strut $^{121}$Sb($\alpha$,$\gamma$)} \hbox{\strut via $\gamma$-ray}}    &   105  & 1  &     \vtop{\hbox{\strut $^{121}$Sb($\alpha$,$\gamma$)} \hbox{\strut via X-rays}} \\

14.00 & 3.64  & 0.08  &    \vtop{\hbox{\strut $^{121}$Sb($\alpha$,n)} \hbox{\strut $^{123}$Sb($\alpha$,n)}}   & 56  & 4 &    \vtop{\hbox{\strut $^{121}$Sb($\alpha$,$\gamma$)} \hbox{\strut  via $\gamma$-ray}}    \\

15.05\footnotemark[3]  & 0.98  & 0.08 &      \vtop{\hbox{\strut $^{121}$Sb($\alpha$,n)} \hbox{\strut $^{123}$Sb($\alpha$,n)}}        \\

16.00& 1.06 & 0.04 &  \vtop{\hbox{\strut $^{121}$Sb($\alpha$,n)} \hbox{\strut $^{123}$Sb($\alpha$,n)}}    \\

\end{tabular}
\end{ruledtabular}
\end{table}

\subsection{$\gamma$-ray counting for the $^{121}$Sb($\alpha$,n)$^{124}$I and $^{123}$Sb($\alpha$,n)$^{126}$I reactions}
The $^{121}$Sb($\alpha$,n)$^{124}$I cross section at each energy from 10.09\,MeV $\leq$ E$_{\alpha}$ $\leq$ 16.00\,MeV was measured by counting the $\gamma$-radiation following the $\beta$-decay of the reaction product. The $^{123}$Sb($\alpha$,n)$^{126}$I cross section was also measured via $\gamma$-ray counting using natural Sb targets at alpha energies between 12.00 and 16.00\,MeV.

The induced activity of the samples was measured using a 100\% relative efficiency high purity Germanium (HPGe) detector in a low background configuration with a commercial 4$\pi$ lead shield. Absolute efficiency calibration of the detector was done at 10 cm and 27 cm distances from the detector crystal, using calibrated $^{7}$Be, $^{22}$Na, $^{54}$Mn, $^{57}$Co, $^{60}$Co, $^{65}$Zn, $^{133}$Ba, $^{137}$Cs, $^{152}$Eu, and $^{241}$Am radioactive sources. At these distances the coincidence summing effect is negligible. 
The left panel of Fig.\,\ref{fig:efficiency} shows the measured efficiency of the HPGe detector. The measured points are fitted by a four-factor function ${\epsilon}$(E) = (AE$^{B}$ + CE$^{D}$)$^{-1}$ \cite{effeq}.
Based on the actual count rate of the reaction products and the dead time of the detector the measurement of the irradiated targets was carried out either at 10 cm or 27 cm from the detector end cap.
In all measurements the data were collected using the ORTEC MAESTRO data acquisition system which provides an automatic dead time control which was measured and found to be precise in an other work \cite{deadtime}.

The ($\alpha$,n) reaction cross sections were measured by counting the yield of $\gamma$-lines listed in the Table\, \ref{tab:decaypar}. To ensure that short-lived activity was minimized, counting typically started 1 -- 4 hours after irradiation had finished. Details of the measurements were summarized in Table\, \ref{tab:meas}. The $\gamma$-spectra were stored at regular intervals (hourly) to follow the decay of the reaction products. The right panel of Fig.\,\ref{fig:spectrum} shows a $\gamma$-spectrum taken for 20 hours after 3 days waiting time on a natural Sb target irradiated with  a 13.07 MeV $\alpha$-beam. The $\gamma$-lines from the decay of $^{121}$Sb($\alpha$,n)$^{124}$I and $^{123}$Sb($\alpha$,n)$^{126}$I reaction products are indicated in Fig.\,\ref{fig:spectrum}.

\subsection{$\gamma$-ray and characteristic X-ray counting for the $^{121}$Sb($\alpha$,$\gamma$)$^{125}$I reaction}
\begin{figure}
\centering
\resizebox{1.1\textwidth}{!}{
\includegraphics[10,10] [1250,450]{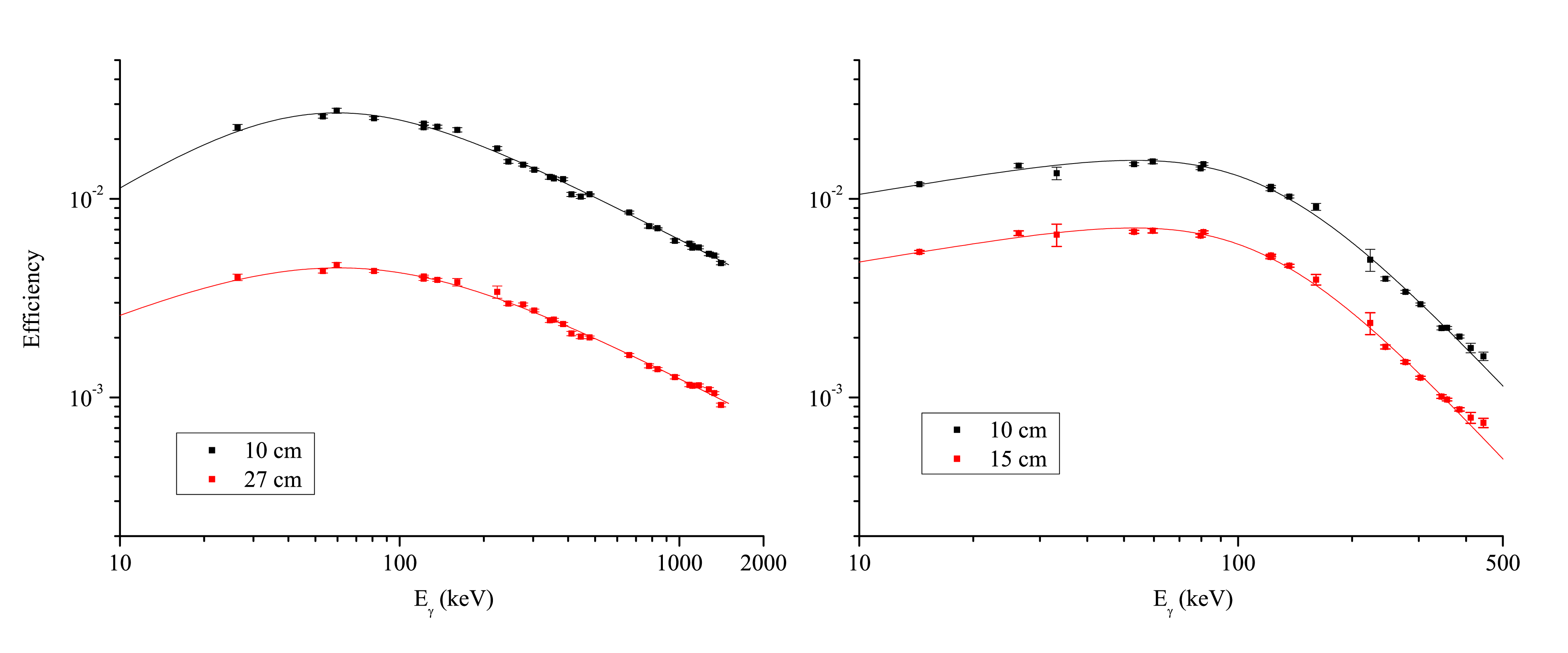}}
\caption{\label{fig:efficiency} (Color online)  Measured detector efficiency of the HPGe detector (left panel) and the LEPS detector (right panel). The obtained efficiency points were fitted by a four-factor function ${\epsilon}$(E) = (AE$^{B}$ + CE$^{D}$)$^{-1}$ \cite{effeq} for both detectors.}
\end{figure}

In order to measure the $\gamma$-ray and characteristic X-rays from the decay of $^{125}$I, a so-called LEPS (Low Energy Photon Spectrometer) detector was used. This type of detector has a thin germanium crystal with large surface and a thin Be entrance window. The LEPS detector was shielded with 8 cm of lead and inner layers of 2 mm cadmium and 4 mm copper \cite{T11}. The absolute efficiency of the LEPS detector was measured with calibrated $^{57}$Co, $^{133}$Ba, $^{152}$Eu and $^{241}$Am sources at 10 cm and 15 cm distances from the crystal of the detector.
The right panel of Fig.\,\ref{fig:efficiency} shows the measured and fitted efficiency of the LEPS detector.
The $^{121}$Sb($\alpha,\gamma$)$^{125}$I cross section was measured by detecting the 35.49 keV $\gamma$-ray and K$_{\alpha 1,2}$ (27.20 and 27.47 keV) characteristic X-rays in the energy ranges of 11.50 MeV $\leq$ E$_{\alpha}$ $\leq$ 14.00 MeV and  10.09 MeV $\leq$ E$_{\alpha}$ $\leq$ 13.50 MeV, respectively. For testing the consistency of these methods, the cross section was measured with both counting methods at 11.50, 12.53 and 13.50 MeV alpha energies. In order to measure the low induced activities, the targets were placed at a position of 3 cm from the LEPS crystal. Because of the strong summing effects at this close geometry, the efficiency measurement was not carried out at this distance directly. Instead, a relative close (3 cm) and far (10\,cm and 15\,cm) geometry efficiency measurement was performed by using the activated Sb targets. For the 35.49\,keV $\gamma$-ray detector efficiency measurements, two natural Sb targets were irradiated at $\alpha$-energies of 15.05 MeV and 16.00 MeV. In addition, for the characteristic K$_{\alpha 1,2}$ X-ray detector efficiency measurements, an enriched $^{121}$Sb target was irradiated at 13.50 MeV. From the counting of these targets at both geometries, a conversion factor for the detector efficiencies at close and far geometries was deduced taking into account the elapsed time between the two measurements.

\begin{figure}
\centering
\resizebox{0.95\textwidth}{!}{
\includegraphics[50,50] [1850,1050]{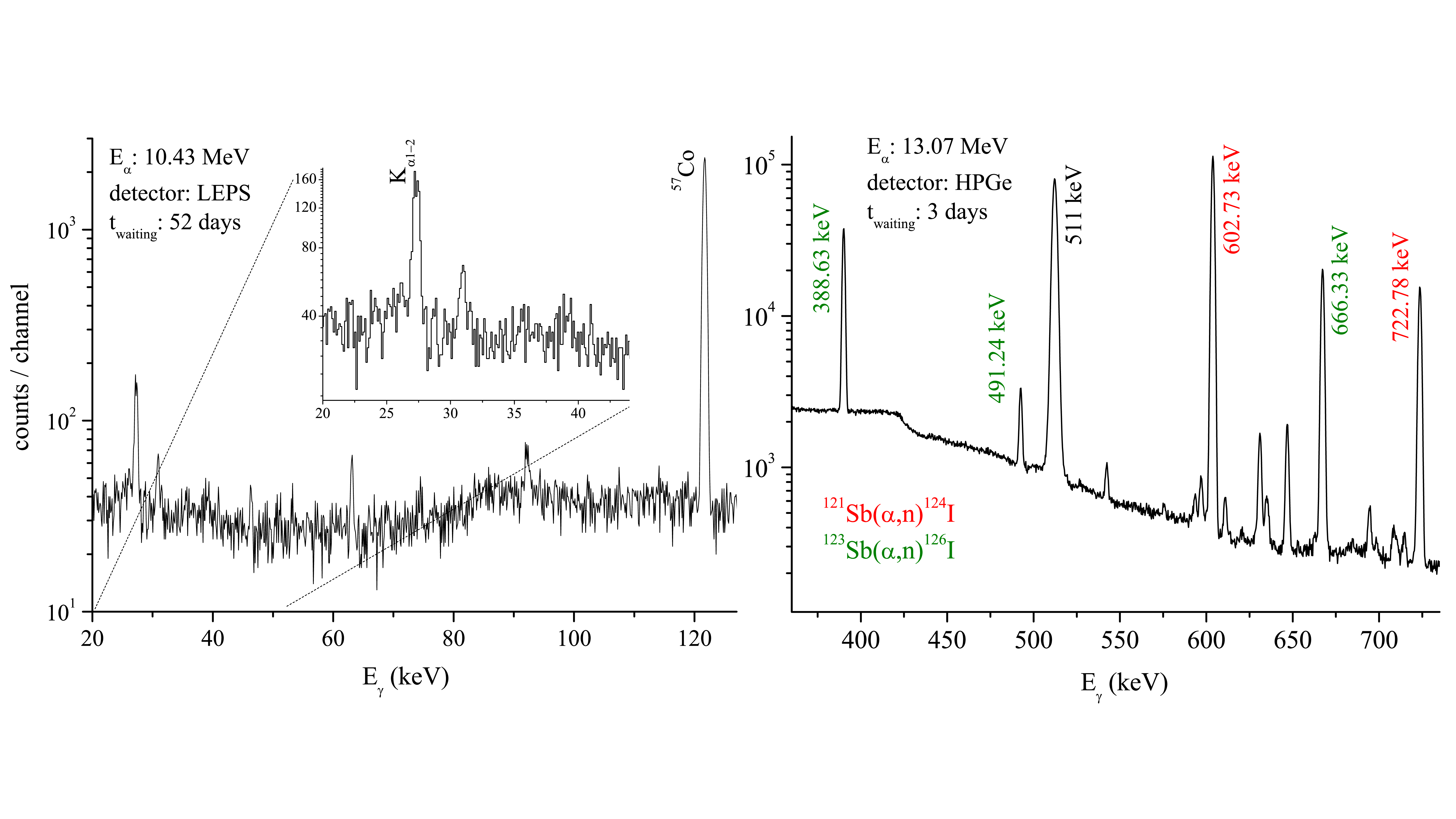}}
\caption{\label{fig:spectrum} (Color online) Characteristic X-ray and a $\gamma$-ray spectra recorded with the LEPS detector (left panel) and the HPGe detector (right panel). The peaks used for analysis are indicated.}
\end{figure}

As discussed in Sec.\,\ref{sec:invesreac} the X-ray counting started at least 73 days after the irradiation (see Table\, \ref{tab:meas}). Consequently the X-ray spectra were dominated by the decay of $^{125}$I.
Counting was performed for 0.3 to 3 days and the spectra were saved every 2 hours. The low energy part of the X-ray spectrum taken on a target irradiated at E$_{\alpha}$ = 10.43\,MeV, is shown in the left panel of  Fig.\,\ref{fig:spectrum}. The resolution of the LEPS detector is between 400\,eV (for a 59\,keV $\gamma$-line) and 680\,eV (for a 122\,keV $\gamma$-line). For the cross section determination, the combined yield of two characteristic K$_{\alpha}$ lines were used because the K$_{\alpha 1}$ and K$_{\alpha 2}$ lines are too close in energy to be resolved (see Fig.\,\ref{fig:spectrum}).

\section{Results and discussion}
\label{sec:results}
\subsection{Measured cross sections}
The $^{121}$Sb($\alpha$,$\gamma$)$^{125}$I, $^{121}$Sb($\alpha$,n)$^{124}$I and $^{123}$Sb($\alpha$,n)$^{126}$I cross sections were measured in the effective center of mass energy ranges of 9.74 -- 13.54\,MeV, 9.74 -- 15.48\,MeV and 11.60 -- 15.48\,MeV, respectively. The measured cross sections for the three reactions are summarized in Table \ref{tab:csres}. The effective center of mass energies were calculated by taking into account the energy loss of the beam in the target layer. The effective energies correspond to beam energies in the target for which one-half of the yield for the full target thickness is obtained \cite{Rolfs88,Ili07}. 

In Table \ref{tab:csres}, the obtained $^{121}$Sb($\alpha$,$\gamma$)$^{125}$I cross sections based on either $\gamma$-ray or X-ray counting are listed separately. 
At energies where both methods were used (11.11, 12.11, and 13.05\,MeV), the cross sections are in good agreement 
as indicated in the fifth and sixth columns of Table \ref{tab:csres}. This consistency check increases the reliability of the determined cross sections. 
At these energies the average cross section values weighted by their statistical uncertainty were adopted as the final results (see Table \ref{tab:csres}). In order to obtain the final uncertainties of the adopted values, the following 
common systematic errors were added quadratically to the statistical uncertainties: current measurement (3\%) and target thickness (4\%).

The uncertainty of the measured cross sections stems from the following partial errors: efficiency of the HPGe detector (6\%) and the LEPS detector (10.9\% for X-rays and 12.4\% for the 35.49 keV $\gamma$-line), the number of target atoms (4\%), measurement of the current (3\%), uncertainty of decay parameters (less than 3\%, see Table\,\ref{tab:decaypar}) and counting statistics (2 -- 14.4\% for the ($\alpha$,$\gamma$) measurements and 0.1 -- 2.2\% for the ($\alpha$,n) measurements). For the X-ray measurements of the ($\alpha$,$\gamma$) reaction, the systematic uncertainty of 1.7\% have been added to the uncertainty of net peak area determination in order to account for a possible contribution of the $^{123}$Sb($\alpha$,n)$^{126}$I reaction to the X-ray peak.

The uncertainties of the center of mass energies given in the second column of Tables \ref{tab:csres} contain the uncertainty of the stopping power of alpha particles in Sb, calculated with the SRIM code (5\%) \cite{SRIM}, Sb target thickness uncertainty (4\%) and the uncertainty of the beam energy of the cyclotron (0.5\%). For the energies (10.09\,MeV, 10.43\,MeV, 12.53\,MeV, 13.07\,MeV and 15.05\,MeV) achieved with degrader foil, the uncertainties arising from energy loss in Al foils (7\%) have also been included.

\begin{table}
\caption{\label{tab:csres} Measured cross sections of the $^{121}$Sb($\alpha$,$\gamma$)$^{125}$I, $^{121}$Sb($\alpha$,n)$^{124}$I and $^{123}$Sb($\alpha$,n)$^{126}$I reactions.}
\setlength{\extrarowheight}{0.1cm}
\begin{ruledtabular}
\begin{tabular}{ccccccc}  
\toprule 
{E$_{\alpha}$}  &  {E$^{eff}_{c.m.}$}  & {$^{121}$Sb($\alpha$,n)$^{124}$I}  & {$^{123}$Sb($\alpha$,n)$^{126}$I} &  & {$^{121}$Sb($\alpha$,$\gamma$)$^{125}$I} &    \\

\cline{5-7}

[MeV] &  [MeV] &   via $\gamma$-ray [mb] & via $\gamma$-ray [mb] & via $\gamma$-ray [$\mu$b] & via X-rays [$\mu$b] & adopted value [$\mu$b] \\

\hline

10.09\footnote[1]{measured with an energy degrader foil.}$^{,}$\footnote[2]{enriched targets; the others are natural targets.}& 9.74 $\pm$ 0.11  & 0.054 $\pm$ 0.004 &   &    & 1.48 $\pm$ 0.22  &  1.48 $\pm$ 0.22   \\

 10.43\footnotemark[1]$^{,}$\footnotemark[2] & 10.08 $\pm$ 0.12 & 0.155 $\pm$ 0.012 & &  & 3.32 $\pm$ 0.43  &  3.32 $\pm$ 0.43  \\

 11.00\footnotemark[2]& 10.62 $\pm$ 0.08 &  0.385 $\pm$ 0.031 &   &  & 5.77 $\pm$ 0.77  &  5.77 $\pm$ 0.77  \\

11.50\footnotemark[2] & 11.11 $\pm$ 0.08 &  1.33 $\pm$ 0.11 &    & 13.3 $\pm$ 1.9 & 12.5 $\pm$ 1.6 &  12.8 $\pm$ 1.3   \\

12.00& 11.60 $\pm$ 0.08 &      2.87 $\pm$ 0.23 & 3.05 $\pm$ 0.25  &24.4 $\pm$ 4.5  &     &    24.4 $\pm$ 4.5      \\

12.53\footnotemark[1]$^{,}$\footnotemark[2]& 12.11 $\pm$ 0.12 &  8.32 $\pm$ 0.66     &   &  45.3 $\pm$ 6.6  & 44.6 $\pm$ 5.6 &    44.9 $\pm$  4.6    \\

 13.07\footnotemark[1] & 12.63 $\pm$ 0.12 &   16.0 $\pm$ 1.3  & 16.6 $\pm$ 1.3 &  70.9 $\pm$ 10.6 &   &  70.9 $\pm$ 10.6   \\

   13.50\footnotemark[2] & 13.05 $\pm$ 0.09 & 32.4 $\pm$ 2.6 &  & 107 $\pm$ 15 & 115 $\pm$ 14 &   111 $\pm$ 11   \\

14.00 & 13.54 $\pm$ 0.10 &  55.5 $\pm$ 4.4 & 58.0 $\pm$ 4.6 &  147 $\pm$ 21 &    &   147 $\pm$ 21  \\

  15.05\footnotemark[1] & 14.55 $\pm$ 0.13   &  189 $\pm$ 15 & 184 $\pm$ 15 &  &   &     \\

   16.00& 15.48 $\pm$ 0.11 &  291 $\pm$ 23 & 230 $\pm$ 19  &    &    &  \\

\end{tabular}
\end{ruledtabular}
\end{table}

\subsection{Comparison with Hauser-Feshbach calculations and previous data}
\label{sec:comparison}
\begin{figure}
         \centering
          \resizebox{0.80\textwidth}{!}{
          \includegraphics[-10,10] [700,450]{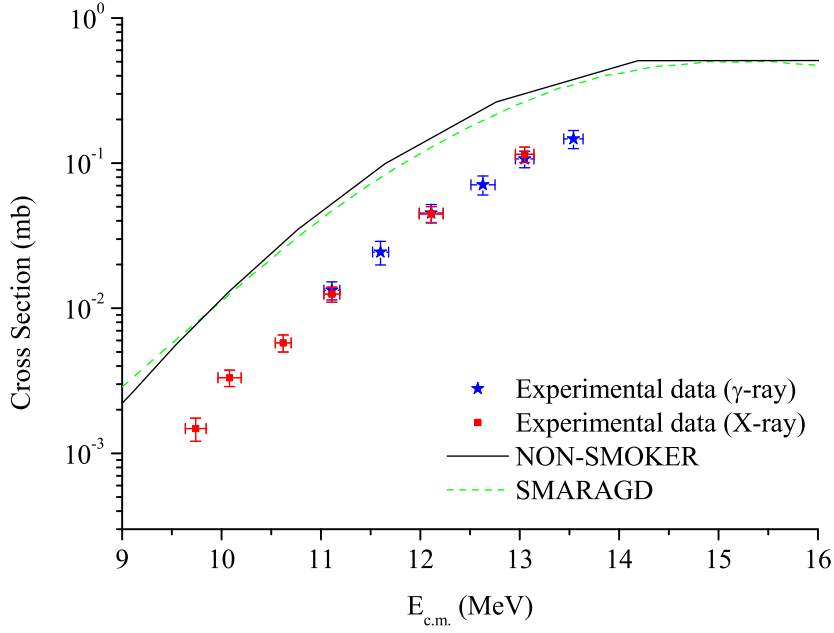}}
         \caption{\label{fig:121agres_Sfac} (Color online) Cross section of the $^{121}$Sb($\alpha$,$\gamma$)$^{125}$I reaction compared with the HF statistical model calculations obtained with standard settings of the statistical model code NON-SMOKER \cite{nuc00,nuc01} (solid line) and SMARAGD \cite{smaragd} (dashed line) with default parameters. The square and star individual forms show the calculated cross section results from X-ray and $\gamma$-ray countings, respectively.}
         \end{figure}

\begin{figure}
         \centering
          \resizebox{0.80\textwidth}{!}{
          \includegraphics[-10,10] [700,450]{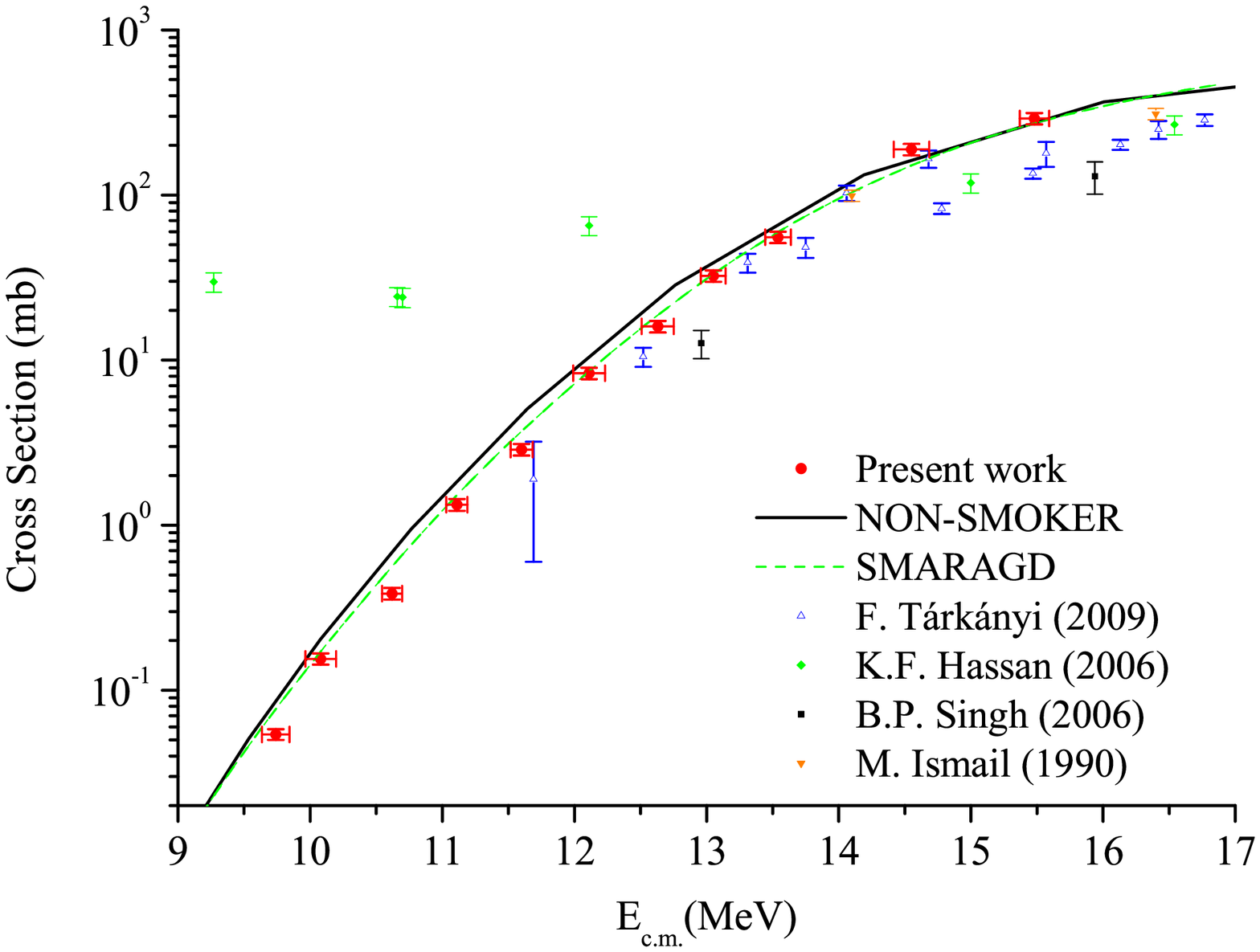}}
         \caption{\label{fig:121anres_Sfac} (Color online) Cross section of the $^{121}$Sb($\alpha$,n)$^{124}$I reaction compared with the HF statistical model calculations obtained with standard settings of the statistical model code NON-SMOKER \cite{nuc00,nuc01} (solid line) and SMARAGD \cite{smaragd} (dashed line) with default parameters. Measured cross sections are also compared with the results of previous experiments \cite{Tark09,Has06,Is90,Sin06}. The preliminary results of the $^{121}$Sb($\alpha$,n)$^{124}$I reaction cross sections were reported in \cite{myproc} with larger energy uncertainties at alpha beam energies of 10.09, 10.43, 
12.53, 13.07, and 15.05 MeV than those in the present paper due to the different accuracy of determining the energy degrader foil thickness. }
         \end{figure}

\begin{figure}
            \centering
             \resizebox{0.80\textwidth}{!}{
             \includegraphics[-10,10] [700,450]{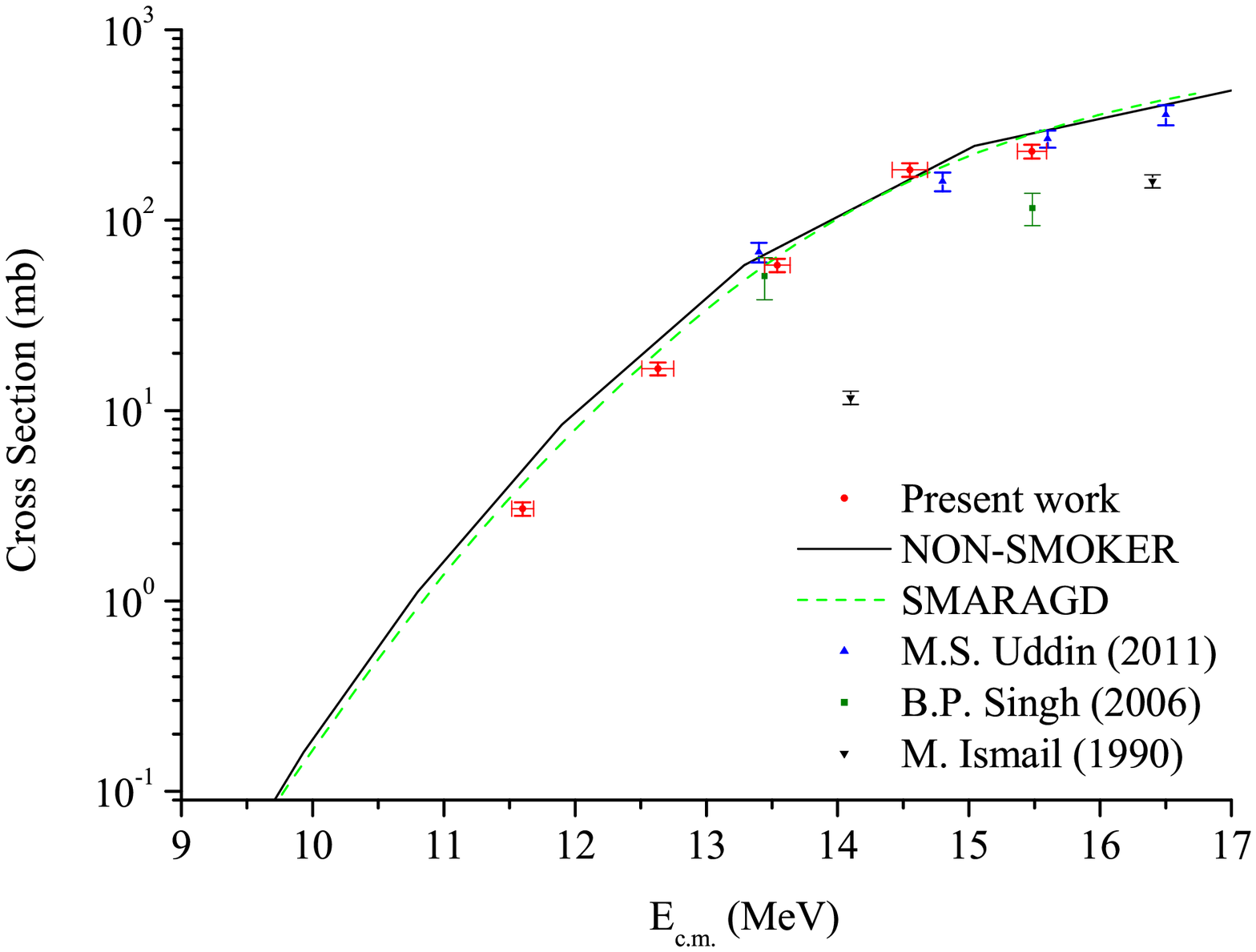}}
            \caption{\label{fig:123anres_Sfac} (Color online) Cross section of the $^{123}$Sb($\alpha$,n)$^{126}$I reaction compared with the HF statistical model calculations obtained with standard settings of the statistical model code NON-SMOKER \cite{nuc00,nuc01} (solid line) and SMARAGD \cite{smaragd} (dashed line) with default parameters. Measured cross sections are also compared with the results of previous experiments \cite{Uddin10,Sin06,Is90}.}
            \end{figure}

The cross section results of all three reactions are shown in Figs.\,\ref{fig:121agres_Sfac}, \ref{fig:121anres_Sfac} and \ref{fig:123anres_Sfac}. The experimental results for $^{121}$Sb($\alpha$,$\gamma$)$^{125}$I are systematically lower by a factor of 2 to 4 compared to the Hauser-Feshbach statistical model calculations of \cite{nuc00,nuc01} (NON-SMOKER) and those obtained with the code SMARAGD \cite{smaragd}.

For the cross section of the $^{121}$Sb($\alpha$,n)$^{124}$I and $^{123}$Sb($\alpha$,n)$^{126}$I reactions some former measurements are available in the literature. For comparison the results of these experiments are also included in Figs.\,\ref{fig:121anres_Sfac} and \ref{fig:123anres_Sfac}. Our present results are closer to the astrophysically relevant energy region and they are in a good agreement with statistical model calculations obtained with standard settings of both codes.

The results for the $^{121}$Sb($\alpha$,n)$^{124}$I reaction given by \cite{Tark09} are scattered and bear an uncertainty of about 40\% at the lowest energy. The data given by \cite{Has06} are systematically lower by a factor of 2 compared to the NON-SMOKER calculations for energies above 14 MeV and rather inconsistent values are reported below this energy. The reaction cross sections at low energies from Ref.\,\cite{Has06} were measured using Al-Sb-Ti sandwich targets resulting in large uncertainties (10 -- 24$\%$) in $\alpha$-particle effective energy at the targets. 
The values from Ref.\,\cite{Sin06} seem to be lower compared to the statistical model calculations, while the two data points at high energies by \cite{Is90} are in good agreement with the statistical model codes.

For the $^{123}$Sb($\alpha$,n)$^{126}$I reactions as seen in Fig.\,\ref{fig:123anres_Sfac}, there is a good agreement between our experimental results and the data taken from Ref.\,\cite{Uddin10} at high energies. The values from Ref.\,\cite{Sin06} and \cite{Is90} for $^{123}$Sb($\alpha$,n) are not consistent with our data and with the model calculations.

\subsection{Astrophysical implications}
\label{sec:astro}

Of interest for the calculation of the astrophysical reaction rate is a comparison of the calculated $\alpha$-width with the one derived from the experimental data. This is because the $\alpha$ width is smaller than the $\gamma$ width in the energy window relevant for the calculation of the rate and thus determines the temperature dependence and absolute magnitude of the rate. Previous investigations found that the $\alpha$ width seems to be strongly overpredicted for intermediate and heavy target nuclei at low energies (but still above the astrophysically relevant energy window), even when there is good agreement with measurements at higher energy (see, e.g., the experimental work cited in Sec.\ \ref{sec:intro}). The picture is not clear, however, since some reactions showed larger discrepancies than others.

\begin{figure}
       \resizebox{0.90\textwidth}{!}{
\includegraphics[angle=0,width=\columnwidth]{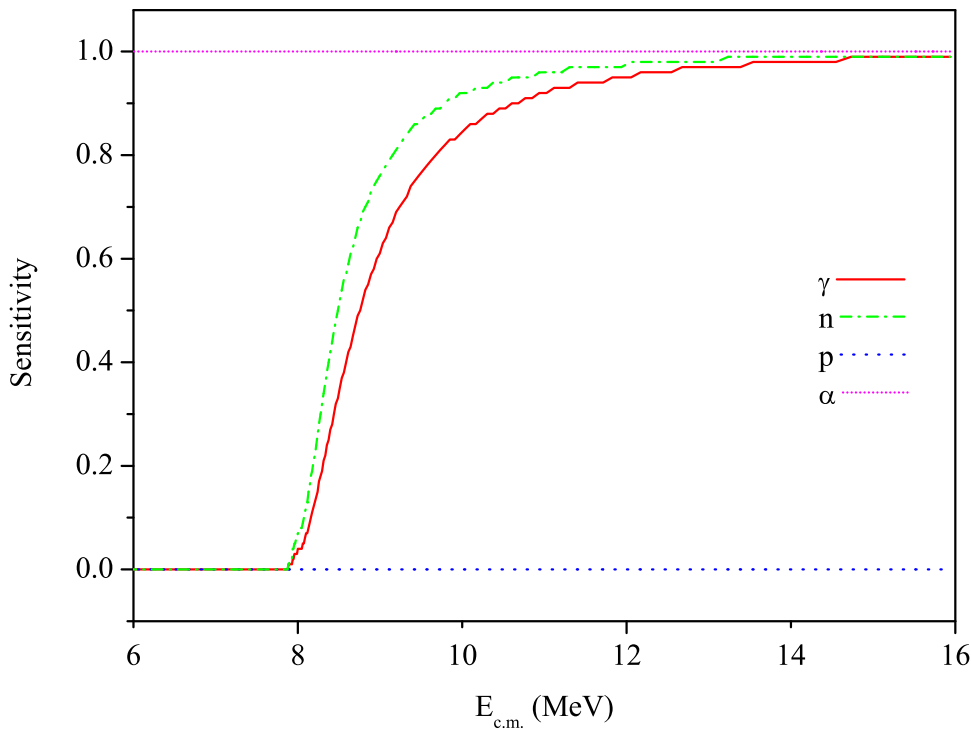}}
\caption{\label{fig:sensi121ag} (Color online) Absolute value of the sensitivity of the reaction cross section of $^{121}$Sb($\alpha$,$\gamma$)$^{125}$I to the variation of particle and radiation widths as function of center-of-mass energy.}
\end{figure}

\begin{figure}
       \resizebox{0.90\textwidth}{!}{
\includegraphics[angle=0,width=\columnwidth]{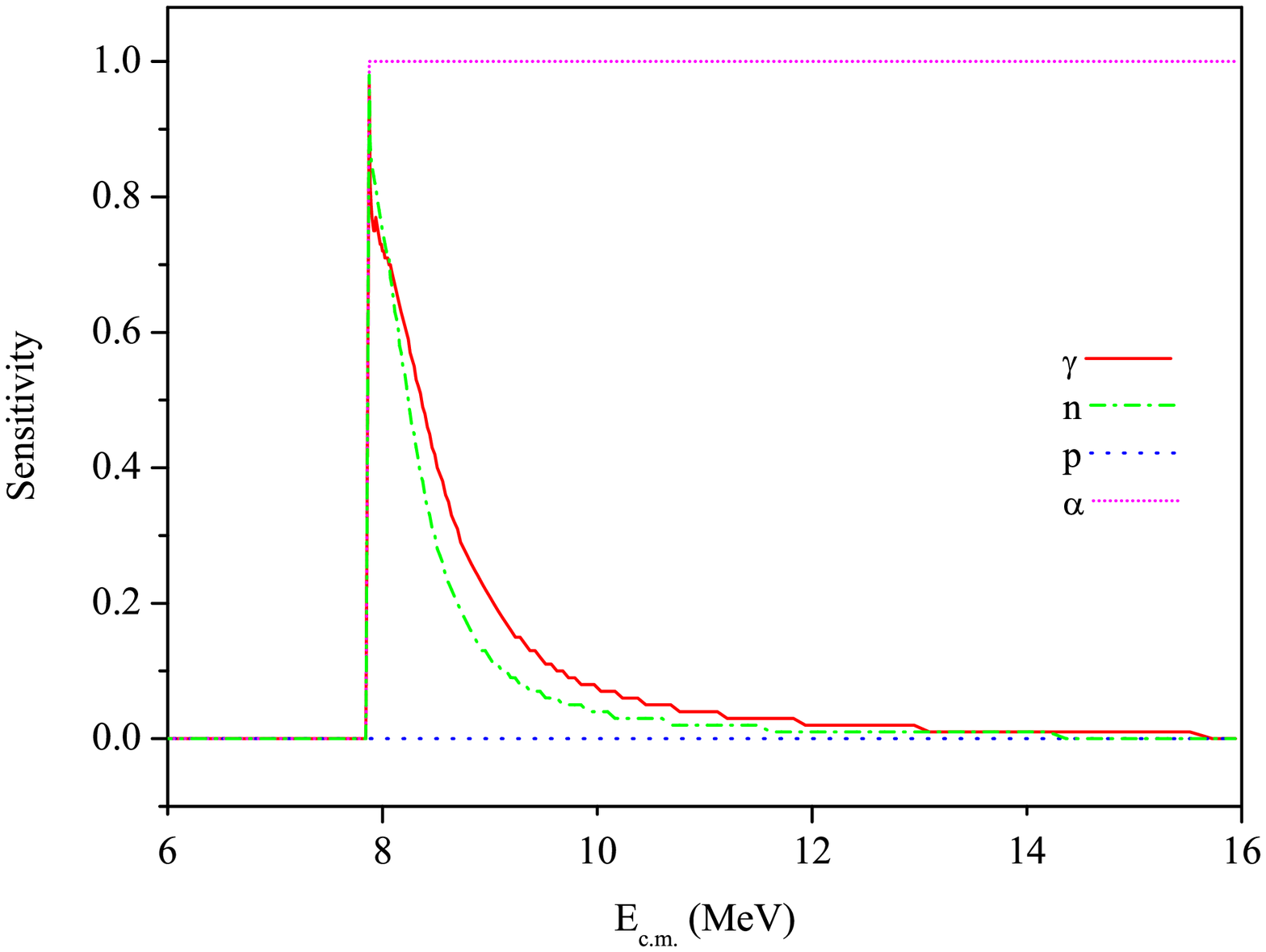}}
\caption{\label{fig:sensi121an} (Color online) Absolute value of the sensitivity of the reaction cross section of $^{121}$Sb($\alpha$,n)$^{124}$I to the variation of particle and radiation widths as function of center-of-mass energy.}
\end{figure}

\begin{figure}
       \resizebox{0.90\textwidth}{!}{
\includegraphics[angle=0,width=\columnwidth]{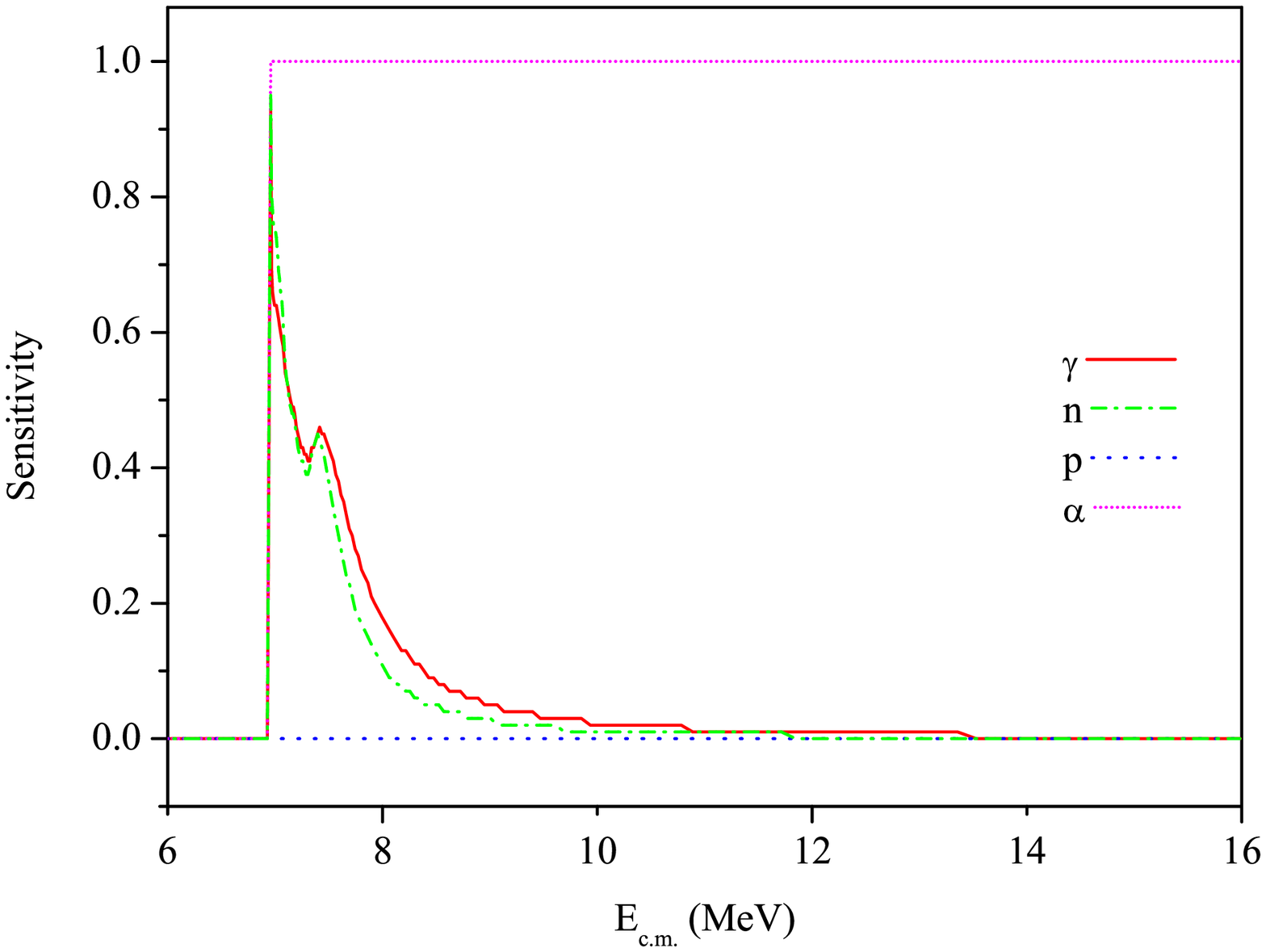}}
\caption{\label{fig:sensi123an} (Color online) Absolute value of the sensitivity of the reaction cross section of $^{123}$Sb($\alpha$,n)$^{126}$I to the variation of particle and radiation widths as function of center-of-mass energy.}
\end{figure}

It is not possible to directly infer the $\alpha$ width from the present $^{121}$Sb($\alpha$,$\gamma$)$^{125}$I data. In the measured energy range, these ($\alpha$,$\gamma$) cross sections depend not only on the $\alpha$ width but also on the neutron and $\gamma$ widths. Only below the ($\alpha$,n) threshold, the ($\alpha$,$\gamma$) cross sections are solely sensitive to the $\alpha$ width. This is illustrated in Fig.\,\ref{fig:sensi121ag}, showing the sensitivities of the reaction cross sections to variations in the widths. A sensitivity of 1.0 implies that the cross section changes by the same factor with which the width is being changed, zero sensitivity means it does not change, according to the definition of the sensitivity given in \cite{sensi}. Discrepancies between predictions and data can be due to mispredictions of either width (or any combination) and therefore it cannot be unambiguously decided by inspection of the ($\alpha$,$\gamma$) cross sections which of the width prediction has to be improved. The ambiguity can be partially lifted by combining the ($\alpha$,$\gamma$) and ($\alpha$,n) data. As shown in Fig.\ \ref{fig:sensi121an} the ($\alpha$,n) cross sections are only depending on the $\alpha$ width, except close to the ($\alpha$,n) threshold. The excellent reproduction of the $^{121}$Sb($\alpha$,n)$^{124}$I data by the NON-SMOKER and SMARAGD predictions shows that the $\alpha$ width is well predicted. This implies that the discrepancies found in comparison to the $^{121}$Sb($\alpha$,$\gamma$)$^{125}$I data have to be attributed to the neutron- and $\gamma$ width. The same is true for $^{123}$Sb+$\alpha$, for which no ($\alpha$,$\gamma$) data have been obtained in this work (the reaction product is stable). The sensitivities for $^{123}$Sb($\alpha$,n)$^{126}$I are shown in Fig.\ \ref{fig:sensi123an}.

This finding adds to the existing database of astrophysical $\alpha$-width studies and underlines the heterogeneity of the overall results. A large difference between experiment and theory has been found for several target nuclei, whereas only small or no differences were found in other cases. This rules out a simple explanation of the phenomenon, e.g., a simple scaling with the Coulomb barrier. Other alternatives have to be explored and the extended data set will aid such investigations. For example, the proposed hypothesis that additional reaction channels, not included in the Hauser-Feshbach calculations, such as low-energy Coulomb excitation \cite{RauPRL}, is consistent with the present data because the Coulomb excitation effect at low energy is negligible for $^{121,123}$Sb due to the spectroscopic properties of these nuclei.


\section{Summary and conclusion}
\label{sec:conclusions}
The cross sections of the reactions $^{121}$Sb($\alpha$,$\gamma$)$^{125}$I, $^{121}$Sb($\alpha$,n)$^{124}$I, and $^{123}$Sb($\alpha$,n)$^{126}$I have been measured with the activation method. The
results are compared with statistical model calculations obtained with the NON-SMOKER and SMARAGD codes. 
The results for $^{121}$Sb($\alpha$,$\gamma$)$^{125}$I represent the first experimental data in the effective center of mass energy range between 9.74 and 13.54 MeV. For energies below the ($\alpha$,n) threshold (8.1 MeV), the ($\alpha$,$\gamma$) reaction 
cross section could not be determined; the expected yield from the reaction was 
lower than the detection limit of the present setup.

The cross sections of the $^{121}$Sb($\alpha$,n)$^{124}$I and $^{123}$Sb($\alpha$,n)$^{126}$I reactions were measured and also compared to previous experimental data. Our results are in a good overall agreement with NON-SMOKER and SMARAGD calculations, but deviate strongly from some previous data.

The agreement of the ($\alpha$,n) data with the predictions implies that the astrophysically relevant $\alpha$ width has been correctly predicted within the measured energy range. To further study the energy dependence of the $\alpha$ width towards the astrophysically relevant energy range, it would be necessary to measure $\alpha$ capture below the ($\alpha$,n) threshold.

\begin{acknowledgments}
This work was supported by the ERASMUS Programme under Lifelong Learning Programme of European Commission, the Scientific and Technological Research Council of Turkey (TUBITAK), Grants No. 109T585 (under the EUROCORES - EuroGENESIS research program), the EU COST Action CA16117 (ChETEC), Kocaeli University BAP (grant 2012/026), NKFIH K120666, K108459 and PD104664. CY acknowledges the support through TUBITAK, under the programme of BIDEB-2219. TR is supported by the UK STFC grant ST/M000958/1 and the ERC Advanced grant GA 321263-FISH. 
Special thanks to Mary Beard who edited the manuscript. 
\end{acknowledgments}


\end{document}